\documentclass{article}
\usepackage[utf8]{inputenc}
\usepackage{longtable}
\usepackage{booktabs}
\usepackage{geometry}
\usepackage{amsmath}
\usepackage{array}
\usepackage{parskip}
\usepackage{graphicx}
\usepackage{tipa} 

\title{Human Voice is Unique}

\author{
  Rita Singh and Bhiksha Raj\\
  Center for Voice Intelligence and Security, Carnegie Mellon University, Pittsburgh, USA
}

\date{}  

\newcommand{\E}{\mathcal{E}}
\newcommand{\M}{\mathcal{M}}
\newcommand{\B}{\mathcal{B}}

\begin{document}

\maketitle

\begin{abstract}
Voice is increasingly being used as a biometric entity in many applications. These range from speaker identification and verification systems to human profiling technologies that attempt to estimate myriad aspects of the speaker's persona from their voice. However, for an entity to be a true biometric identifier, it must be unique. This paper establishes a first framework for calculating the uniqueness of human voice objectively. The approach in this paper is based on statistical considerations that take into account a set of measurable characteristics of the voice signal that bear a causal relationship to the vocal production process, but are not inter-dependent or derivable from each other. Depending on how we quantize these variables, we show that the chances of two people having the same voice in a world populated by 10 billion people range from one in a few thousand, to one in a septillion or less. The paper also discusses the implications of these calculations on the choices made in voice processing applications.
 \end{abstract}

\section{Introduction}


While it has been conjectured that fingerprints and DNA are unique, the uniqueness of human voice has never been in contention. Unfortunately, this has been so for the wrong reasons: perceptually, it is difficult for most people to believe that human voice is unique. Many voices that we hear in the course of our lifetimes sound similar (or the same) to us. This perception is exacerbated by the fact that we rarely hear the voices we consider to be the same, together. In reality our \textit{memory} of acoustic phenomena in general, and of speech sounds in particular, plays a dominant role in the comparisons we make.

This brings the clarity of our perception in focus -- could our perception be biased by the sequentiality of our auditory observations? Could it be wrong? The fact that we perceive many voices to be similar could also be a consequence of the shortcomings of our auditory system -- our auditory system may be limited in its sensory capacity, or our brains may have become conditioned to ignore some of the most discriminating nuances of human voices for evolutionary, biological or other reasons.  

When we disregard our perceptual biases, and search for more clarity in this context through more objective criteria, it is reasonable to examine the process of voice production in humans first, to understand the acoustic nature of voice. When we do so, the first thing that  strikes us is its complexity -- the large number of variables involved in it, and the role played in their formation and function by factors that are both a) highly variable and dynamic (such as the physical and socio-cultural environment), and b) highly unique to individuals (such as gene expression and DNA). It strikes us that such vast numbers of intricate variables cannot all be the same for two people -- in fact the latter begins to strike us as a virtual impossibility. Is human voice then unique? Can we objectively prove or disprove a hypothesis that human voice is unique? The goal of this paper is to answer these questions.

Some facets of vocal uniqueness in humans have been addressed in the scientific literature indirectly. There has been some prior work that has focused on either \textit{Acoustic uniqueness}: demonstrating statistically significant speaker differences in controlled corpora e.g. \cite{Kinnunen2010}; or \textit{Perceptual confusability}: estimating how often listeners mistake one speaker for another in limited language contexts, e.g. \cite{Jessen2022}. Neither line of inquiry delivers an \emph{absolute} probability that two arbitrary humans could, by chance, share an indistinguishable voice.

In this paper, we examine the hypothesis that human voice is unique from an objective and statistical perspective. From this perspective, the first line of reasoning about the uniqueness of the human voice naturally comes from a consideration of the specific aspects of the voice production process that are different from speaker to speaker. 

The process of voice production in humans has been studied and modeled for over a century. It is estimated that well over a hundred thousand scientific papers that document the results of myriad studies in this context have been published in the recorded literature so far. The complexity of the human voice production system is well-understood \cite{Singh2019production}. The physical, physiological and cognitive factors that drive it are well known. Yet, the uniqueness of human voice has never been established. In fact, to the contrary, there have been many attempts across many fields of study toward categorizing human voices into a small, human-manageable number of ``types''  based on their perceived commonalities.  For example, in the field of performing arts, specifically Opera, voices have been categorized into 7 types and even more subtypes: Soprano, Mezzo-soprano, Contralto, Tenor, Baritone, Bass and Countertenor, with extensions to Treble, Castrato, Haute-contre and the Fachs subcategories (which can number over 25). These categorizations are based on the singing quality of voices \cite{McGinnis2010}.

In contrast, medical fields such as otorhinolaryngology and psychology employ subjective categorizations based on entities called ``voice qualities,'' which capture the perceptual ``flavor'' or ``sound'' of a voice (e.g. wobbly, breathy, nasal, rough, strained, pressed etc.) \cite{Aronson1985}  In otorhinolaryngology, clinical studies use these qualities to assess vocal health, often through standardized scales like the GRBAS (Grade, Roughness, Breathiness, Asthenia, Strain) scale \cite{Hirano1981}. Similarly, in psychology and psychiatry, subjective voice quality assessments, such as those in the CAPE-V (Consensus Auditory-Perceptual Evaluation of Voice) scale \cite{Kempster2009}, serve as diagnostic aids to evaluate emotional or psychological states reflected in vocal characteristics. These scales rely on trained listeners' perceptual judgments, complementing objective measures like acoustic analysis, and are critical for diagnosing voice disorders or mental health conditions.

The fact that voice carries biometric information about the user has been widely recognized and documented. It is interesting to note that the categorization of human voices into a small number of categories based on subjective and perceptual criteria \cite{Ball1995, Garellek2022, Bhuta2004} has also constrained the attempts to categorize the biometric information derivable from voice into a small number of categories, and also sought to find broad commonalities in objectively measured acoustic properties that correspond to these categories \cite{Abercrombie1967, Barsties2015}. This has clearly driven many assumptions made in studies that have attempted to correlate the biometric information to the specific (subjective) voice categories mentioned above -- often a few limited voice-quality classes are used when modeling or evaluating speaker traits, or assessing speaker identity \cite{Kuwabara2015, Podesva2015}.
 
In this paper, we draw away from the perceptually-based categorizations mentioned above to bring an alternate ``uniqueness hypothesis'' to the forefront: we hypothesize that \textit{given the complexity of the human vocal production process, human voice is likely to be unique}. We note that this hypothesis is partially and implicitly supported by the recent success of data-driven voice-based identification and verification systems that are successfully able to discriminate between increasingly large numbers of speakers \cite{Ohi2021,Jahangir2021}. If voices were not unique, and only a small number of distinct voice ``types'' existed, the performance of such systems would be limited to well below currently reported levels of accuracy.

In the sections below, we present our approach to selecting and using objective evidence to calculate the uniqueness of human voice.

\section{Selecting the bases for uniqueness calculations}

To support our approach,  some facts about the human vocal production system must first  be reviewed.

Voice occupies a singular niche among all biometric identifiers in that it is simultaneously driven by a bio-mechanical process that is tied to the physical and physiological condition of the speaker, and continuously molded by a cognitive process that is conditioned by physiological, psychological, cultural, environmental and other influences. The bio-mechanical process of voice production begins with the generation of sub-glottal pressure from the lungs. This, through a well-studied chain of physical forces, causes self-sustained vibratory motion of the vocal folds. This motion modulates the glottal air stream, producing a pressure wave that undergoes further changes within the supra-laryngeal region. Even very small changes in the vocal-fold length and tissue stiffness, epilaryngeal area, palate arching, mandibular angle, or total vocal-tract length \cite{Fitch1997}, among other things, shift the resulting \textit{spectral and temporal patterns} in the voice signal produced in ways that listeners perceive as ``different'' voices \cite{Titze1994, Behrman2021, Singh2019profiling}.

Shifts in the spectral and temporal patterns of the voice signal are in fact caused by any perturbation -- large or small -- within the entire voice production process. As examples, we expand on the role of anatomical and neuro-cognitive factors, noting that these are not the only factors -- many others are known to influence the spectro-temporal patterns in the voice signal.

From the perspective of anatomical variability, three broad classes of anatomical factors drive baseline vocal diversity.
The first class is that of genetic determinants. Allelic variation in genes involved in laryngeal and cranio-facial morphogenesis (e.g.\ \textit{TBX1} \cite{Kelly2004}, \textit{FOXP2} \cite{Xu2018}) sets initial bounds on cartilage dimensions, muscle fiber composition, and tract length.  Heritability studies report a genetic contribution to a speaker's normal pitch (habitual $F_0$) \cite{Przybyla1992, Gisladottir2023}, which also correlates with  the heritability of singing ability (which is closely related to $F_0$) at $\sim$40\%. \cite{Yeom2022}. 

The second class of anatomical factors that drive vocal diversity is skeletal architecture.  Population-wide CT and MRI surveys document systematic differences in skull base angle, maxillary depth, and pharyngeal cavity volume \cite{Raveendranath2022, Wagner2017, Vorperian2009}.  These bony landmarks constrain the articulatory workspace and therefore the attainable formant trajectories \cite{StoryTitze1998, Stevens1998}. Since formants span the resonant frequencies of the vocal tract, these factors affect the perceived ``sound'' of an individual's voice. Small changes in skeletal structure can lead to perceivable differences in the sound of people's voices, an inference that is also supported by direct studies, e.g. \cite{Macari2016}.

The third important set of anatomical factors is soft-tissue growth and aging.  To mention a few examples from the many recorded changes in this category: vocal-fold lamina propria layers thicken and lose elasticity with age, altering both mean $F_0$ and vibratory aperiodicity (jitter/shimmer) \cite{Sato1997}.  Concurrent descent of the larynx lengthens the pharyngeal cavity, lowering formant frequencies.

Anatomy alone, however, does not account for all variability in voice.\cite{Cavalcanti2021}. An equally important role is that of neuro-cognitive modulation. Cortico-bulbar pathways fine-tune motor commands that set sub-glottal pressure, vocal-fold tension, articulator positions, and prosodic patterns in real time \cite{Juergens2002}.  Language-specific phonotactics, sociolinguistic identity, emotional state, health and myriad other factors further modulate these commands on millisecond time scales.  Long-term learning shapes articulatory habits so effectively that identical twins raised apart exhibit measurably divergent vowel spaces and intonation styles despite near-identical laryngeal morphologies \cite{Mayr2012}.

\subsection{Using spectral and spectro-temporal measurements as the bases}

From the brief account of the sources of variability above, we see that many factors that influence the voice production process are reflected in the spectral and spectro-temporal characteristics of the speech signal. It is therefore reasonable to base our considerations of uniqueness on these characteristics (or ``features''). From a survey of existing literature, we identify the strongest features from the perspective of their relationship and responsiveness to the vocal production process, and their causal independence from each other. Our survey identifies 44 such features, which we use as the basis for further calculations. The following section gives an account of these features.

\section{Feature selection}

Table \ref{tab:independent-features} lists the features we select for our estimates, after a broad survey of the literature. These are all well-known and well-described in the literature, and represent measurable characteristics of the acoustic signal whose values arise from distinct, primary aspects of voice production, while being causally independent of each other. ``Independent'' here means the value is a primary control or physical parameter in the speech production chain. Other widely-used characteristics (outside of this list)  may be calculable from some combination of these, but not vice versa. ``Independent from each other'' means that each feature listed in Table \ref{tab:independent-features} taps a distinct physical or physiological control that is not uniquely recoverable from the other features listed in this table. 

Only brief descriptions are provided in Table \ref{tab:independent-features}. Since these features are well-established in the literature, a single supporting reference is provided with each description as a lead for further details.


{\small
\begin{longtable}{|>{\raggedright\arraybackslash}p{0.8cm}|
                  >{\raggedright\arraybackslash}p{3.3cm}|
                  >{\raggedright\arraybackslash}p{9.9cm}|}
\caption{Measurable acoustic features and their causal independence} \label{tab:independent-features} \\
\hline
\textbf{No.} & \textbf{Feature} & \textbf{Causal independence explanation} \\
\hline
\endfirsthead

\hline
\textbf{No.} & \textbf{Feature} & \textbf{Causal independence explanation} \\
\hline
\endhead

\hline
\endfoot

\hline
\endlastfoot
1 & Alpha Ratio (1--5 kHz/50--1 kHz energy) & Widely used brightness metric tied to laryngeal tilt and articulation; unlike spectral tilt (a regression slope) or roll off (single cut off), it is a band energy ratio independent of moment statistics. \cite{SundbergNordenberg2006} \\
2 & Amplitude Modulation Depth (AMD) & Strength of slow ($<$20 Hz) envelope fluctuations from respiratory and laryngeal gestures; cannot be recovered from RMS or modulation spectrum peak alone. \cite{Joris2004} \\
3 & Breath Group Duration (BGD) & Time between successive inspiratory pauses; controlled by respiratory planning and separate from syllabic speech rate counts. \cite{Winkworth1995} \\
4 &	Closed Quotient (CQ)	& Defined as the proportion of the glottal cycle during which the vocal folds are closed. Unlike GCT or SQ, it directly captures vocal fold adduction pattern and impacts voice quality (e.g. pressed voice), but is not predictable from shimmer, jitter, or HNR. \cite{Verdolini1998} \\
5 & $\Delta$ CPP (frame wise change in Cepstral Peak Prominence) & Tracks stability of source--filter coupling across time; complements absolute CPP by isolating temporal variability not conveyed by jitter, shimmer, or spectral flux. \cite{Fraile2014} \\
6-10 & Formant Bandwidths (B1--B5) & Reflect damping/Q of each resonance, governed by tissue and tract wall losses; not predictable from formant frequency alone. \cite{Fant1971} \\
11 & Formant Dispersion & Average spacing between successive formants, governed primarily by vocal tract length; independent of their absolute frequencies or bandwidths already listed. \cite{Fitch1997} \\
12-16& Formant Frequencies (F1--F5) & Set by vocal tract shape \& modulated by articulation; all spectral shape indices (centroid, tilt, bandwidth, Hammarberg, etc.) are consequences of these resonances. \cite{Stevens1998} \\
17 & Fundamental Frequency (F0) & Primary rate of vocal fold vibration. \cite{Titze1994} \\
18 & Glottal Closure Time (GCT) & Fraction of each glottal cycle in which the folds are closed; influences harmonic structure and Harmonic-to-Noise Ratio (HNR) but is not computable from those outcomes alone. \cite{Titze1994} \\
19 & Glottal to Noise Excitation (GNE) & Compares periodic excitation energy to turbulent noise in the 0.3--4 kHz band; reflects degree of incomplete vocal fold closure and supraglottic turbulence, which neither HNR nor SPI nor breathiness indices isolate. \cite{Michaelis1997} \\
20 & Harmonic Richness Factor (HRF) & Sum of harmonic amplitudes relative to F0 amplitude; distinct from HNR (noise vs harmonic) and from spectral tilt. \cite{Childers1991} \\
21 & Inharmonicity Index (IHI) & Average deviation of harmonic frequencies from integer multiples of F0 due to stiffness or asymmetry; not captured by jitter (period) or SHR (subharmonics). \cite{Fletcher2012} \\
22 & Jitter & Captures cycle to cycle variation in F0 period; cannot be derived from any spectral shape or amplitude measure without first measuring the period itself. \cite{Titze1995} \\
23 & Low to High Energy Ratio (LHR) & Energy below 1 kHz divided by energy above 3 kHz; captures overall spectral balance not implied by formant positions or tilt. \cite{Awan2010} \\
24 & Maximum Flow Declination Rate (MFDR) & Rate at which glottal airflow shuts off within each cycle; governed by collision dynamics of the folds and not deducible from F0, jitter, RMS, or GCT. \cite{Titze2006} \\
25 & Nasality Index & Set by velopharyngeal port coupling (oral vs. nasal tract); affects formants and spectral tilt yet is an independent articulatory state. \cite{Chen1997} \\
26 & Normalized Amplitude Quotient (NAQ) & Ratio of peak to peak glottal airflow to MFDR, capturing overall pulse ``openness''; depends on absolute flow shape rather than on timing parameters like SQ or GCT. \cite{Alku2002} \\
27 & Pitch Strength (autocorrelation peak) & Measures degree of periodicity or harmonic salience; orthogonal to mean F0 (rate) and jitter (cycle variability). \cite{Yost1996} \\
28 & Root Mean Square (RMS) Amplitude & Direct measure of acoustic power generated at the glottal source. \cite{Alku1998}
\\
29 & Semitone SD of F0 & Standard deviation of fundamental frequency in semitones across an utterance; a prosodic planning variable independent of mean F0. \cite{Baken1996} \\
30 & Shimmer & Captures cycle to cycle variation in amplitude; distinct from RMS/energy and not inferable from any other single metric. \cite{Titze1995} \\
31 & Sibilant Spectral Peak Frequency (SSPF) & Center frequency of /s/ or /\textipa{S}/ fricative noise, set by oral constriction length; articulatory and unrelated to vowel formants, F0, or energy measures. \cite{Jongman2000} \\
32 & Soft Phonation Index (SPI) & Low vs high frequency energy ratio indicating incomplete glottal closure (breathiness); RMS/tilt/HNR do not uniquely specify it. \cite{Hillenbrand1996} \\
33 & Spectral Entropy & Shannon entropy of the magnitude spectrum; quantifies distribution uniformity and cannot be inferred from any individual spectral moment (centroid, bandwidth, skewness, kurtosis). \cite{Toledano2010} \\
34 & Spectral Flux & Frame to frame Euclidean change in the short time spectrum; captures dynamic spectral movement beyond what modulation spectrum peaks or AMD describe. \cite{Scheirer1997} \\
35 & Spectral Kurtosis & Fourth spectral moment (peakedness/tailedness); provides information orthogonal to centroid, bandwidth, and skewness. \cite{Peeters2004} \\
36 & Spectral Roll Off (95 \% energy) & Frequency below which 95 \% of total energy lies; independent of centroid or bandwidth because they describe different moments. \cite{Scheirer1997} \\
37 & Spectral Skewness & Third spectral moment (asymmetry of the spectrum); not predictable from centroid (first moment) or flatness/tilt metrics. \cite{Peeters2004} \\
38 & Speech Rate & High level motor planning parameter (syllables per second); phoneme durations and modulation spectra follow from it, not the other way round. \cite{GoldmanEisler1972} \\
39 & Speed Quotient (SQ) & Ratio of glottal opening to glottal closing time; a separate temporal shape control unrelated to MFDR or GCT values. \cite{Holmberg1988} \\
40 & Subharmonic to Harmonic Ratio (SHR) & Measures nonlinear vocal fold oscillations that create subharmonics; cannot be inferred from HNR, jitter, or shimmer. \cite{Sun2002} \\
41 & Temporal Fine Structure Phase Coherence & Frame wise consistency of harmonic phases; reflects source periodicity and is not derivable from amplitude only measures. \cite{McAulay1986} \\
42 & Vocal Fry Index (VFI) & Proportion of frames exhibiting creaky voice pulse clustering; independent of jitter/shimmer because it reflects a different vibratory regime. \cite{Wolk2012} \\
43 &	Vocal Tract Length Estimation (VTLE)	& Derived from formant dispersion and resonant spacing but reflects a distinct physical dimension: the absolute length of the speaker's tract. It governs global resonance structure and is independent of articulation, prosody, or laryngeal behavior. \cite{Fitch1999} \\
44 & Voice Onset Time (VOT) & Determined by articulatory timing of stop release vs. voicing; no other metric in this list uniquely specifies this temporal gap. \cite{LiskerAbramson1964} \\
\end{longtable}
}


\section{The collision of voices}

Our objective is to derive an estimate of the probability that two humans have the same voice. For the purposes of our computation, we will consider the 44 acoustic features described above to completely describe the voice\footnote{In reality there may be yet other features not considered here, so all estimates must be viewed as approximations.} Thus, from our perspective, a voice may be viewed as a 44-component ``acoustic feature'' vector, each component of which represents one of these features.  

Two voices are the same if they have the same acoustic feature vector. Since each of the features is continuous-valued, the probability that two randomly drawn individuals will have the same voice by this definition is zero, simply as a property of continuous-valued random variables. More realistically, we must not require exact match, but also consider \textit{tolerances} -- two voices may be considered the same, i.e. to \textit{match}, if their acoustic feature vectors are sufficiently close. We must compute voice-sameness probability in terms of this notion of closeness.

We take a combinatorial approach to computing match probabilities by quantizing the space of acoustic features into $m$ distinct cells, where each cell represents one possible variant of voice. Each of the 44 independent measurable features that can very in a voice signal, is known to vary within a range of continuous values. The boundaries of these ranges have not been mapped across the human population at the time of writing this paper. Nevertheless, we assume that we can divide each of the continuous-valued ranges into $q$ quantized intervals. The number of quantization levels $q$ could be related to perceptual granularity, or in a computational setting, to the resolution of the features -- the larger the $q$ the more finely resolved the features are. Thus, the total number of voice variants is $m = q^{44}$.  Two voices are considered a match if they fall into the same cell, i.e. if they ``collide''.  We must quantify the probability of such collisions. 

We use the following four metrics to quantify this probability:
\begin{enumerate}
\item \textbf{Exact match:} \textit{Given} an individual, this is the probability of finding at least one other individual in the population whose voice matches them. We will refer to exact match by the symbol $\E$ and its probability as $P(\E)$ in the rest of this paper.

\item \textbf{Match with $p$:} In this test, given an individual, we compute the size $S$ of the \textit{smallest} group of randomly drawn other individuals such that the probability that at least one of them matches the given individual is at least $p$. 

\item \textbf{Pair match:} This is the probability that a randomly chosen pair of individuals from the population will have matching voices. We will refer to pair match by the symbol $\M$ and its probability as $P(\M)$ in the rest of this paper.
\item \textbf{Population match:} This is the probability that the entire population includes at least one pair of voices that match. We will refer to population match by the symbol $\B$ (after the \textit{birthday problem}) and its probability as $P(\B)$ in the rest of this paper.
\end{enumerate}
The exact value of these probabilities will depend on both, the number of quantization levels $q$, and the actual (human) population size, which, for the purposes of this paper, we will generously assume to be 10 billion, i.e. $10^{10}$. We note that while exact match and pair match probabilities will generally align with intuition, population match probabilities, which relate to the well-known \textit{birthday problem} \cite{borja2007birthday}, may sometimes surprise us.

We derive each of these probabilities below. In all cases, we set up the problem as follows. A population of $n$ individuals (where, as mentioned above, $n=10^{10}$) is distributed over $m$ cells (where $m = q^{44}$ as mentioned). We assume that the $n$ individuals are IID, i.e. that they are independent, and are assigned to cells according to identical probability distributions. Furthermore, since each of the 44 features are independent, and the $q$ quantization bins for each feature can be chosen to be equiprobable, we assume that this distribution is uniform, in other words, all $m$ cells are equiprobable, with probability $\frac{1}{m}$.

A match is a collision where two individuals end up in the same cell.

We introduce some terminology here to simplify our explanations. In some of the metrics below, we discuss the probability of a \textit{specific} individual being matched by one or more \textit{other} individuals in the population. In order to distinguish the two, we will refer to the specific individual being considered as a \textit{target}, and the remaining individuals as \textit{peers}.

\subsection{Exact match}
An \textit{exact match} is the event $\E$ where, given any target individual $a$, at least one other peer in the population has a voice that falls in the same voice cell as $a$.

Let $a$ be assigned to any cell $c$.  To compute the probability of an exact match to $a$ among the peers, we first compute the complement, i.e. that \textit{no} peer has been placed in $c$.

Since $P(c) = \frac{1}{m}$, the probability that any peer will draw one of the $n-1$ cells that is \textit{not} $c$ is 
\[
P(\bar{c}) = 1 - \frac{1}{m}
\]

The probability $P(\bar{\E})$ that \textit{all} $n-1$ peers select a cell that is not $c$ is thus
\[
P(\bar{\E}) = \left(1 - \frac{1}{m}\right)^{n-1}
\]

The probabilty of an exact match between target $a$ and at least one of the peers is $1 - P(\bar{\E})$, which is given by
\begin{equation}\label{eq:exact0}
P(\E) = 1 - \left(1 - \frac{1}{m}\right)^{n-1}
\end{equation}

For large $m$ Equation \ref{eq:exact0} can be further approximated. We know that for small $x$ (i.e. $x \ll 1$),  $(1 - x)^N \approx 1-Nx$. Using this approximation, we obtain
\[
P(\bar{\E}) = \left(1  -\frac{1}{m}\right)^{n-1} \approx 1 - \frac{n-1}{m}
\]
giving us
\begin{equation}\label{eq:exact01}
P(\E) = 1 - \left(1 - \frac{n-1}{m}\right) = \frac{n-1}{m}
\end{equation}

A better approximation is obtain from $e^{-x} \approx 1 -x,~\text{for}~x \ll 1$.  Applying this approximation to $P(\bar{\E})$ we get

\[
P(\bar{\E}) = \left(e^{-\frac{1}{m}}\right)^{n-1} = e^{-\frac{n-1}{m}}
\]
giving us
\begin{equation}\label{eq:exact1}
P(\E) = 1 - e^{-\frac{n-1}{m}}
\end{equation}

A simple interpretation of $P(\E)$ is that, for any target individual, we may expect to have to (repeatedly) redistribute the remaining $n-1$ peers over the $m$ cells 
\(
N(\E) = \frac{1}{P(\E)}
\)
times before we find an arrangement in which we find a match for the target among them. The larger $N(\E)$, the greater the uniqueness of the target voice.

\subsection{Match with $p$}
The exact match metric merely quantifies how rare it is to find an exact match for a  target individual $a$ within our given population of size $n$, but does not actually inform us how large the population \textit{should} be to find a match with sufficiently high probability (note that the $N(\E)$ described above does not give us this number).  This metric is quantified by \textit{match at $p$}.

\textit{Match at $p$} is the size $S$ of the \textit{smallest} group $G$  of randomly drawn peers such that the probability that at least one of them matches a target  $a$ is at least $p$.  The math derives directly from that of exact match.

For a peer set $G$ of size $S$, the probability that at least one of them will match $a$ is obtained from Equation \ref{eq:exact0} as
\[
P(match~in~G) = 1 - \left(1 - \frac{1}{m}\right)^{S}
\]
We want this to be at least $p$.
\[
1 - \left(1 - \frac{1}{m}\right)^{S} \ge p ~\Rightarrow ~\left(1 - \frac{1}{m}\right)^{S} \le 1-p
\]
Taking logarithms and solving for $S$:
\begin{align}
S \log \left(1 - \frac{1}{m}\right) &\le \log(1-p) \label{eq:p0}\\
S &\ge \frac{\log(1-p)}{\log\left(1 - \frac{1}{m}\right)} \label{eq:p1}
\end{align}
Note that the directionality of the inequality changes between Equations \ref{eq:p0} and \ref{eq:p1} because $\log(1-p)$ and $\log(1-\frac{1}{m})$ are both negative.

Since $S$ must be an integer, the formula is:
\begin{equation}\label{eq:atp0}
S = \left\lceil \frac{\log(1-p)}{\log\left(1 - \frac{1}{m}\right)} \right\rceil
\end{equation}
where $\lceil \cdot \rceil$ represents the ceiling function.

When $m$ is large, this can be further approximated. Using the same approximation $\log(1-x) \approx -x, ~x \ll 1$ (which is just the logarithmic variant of the approximation $e^{-x} \approx 1-x,~ x \ll 1$ we used earlier), we get, for large $m$,
\[
\log\left(1 - \frac{1}{m}\right) \approx - \frac{1}{m}
\]
leading us to
\begin{equation}\label{eq:atp1}
S \approx \left\lceil -m \log(1-p) \right\rceil
\end{equation}

For small $p$, we can, again, use the approximation $\log(1-x) \approx -x, ~x \ll 1$ to get.
\begin{equation}\label{eq:atp2}
S \approx \left\lceil mp \right\rceil
\end{equation}

For any $p$, larger $S$ implies greater uniqueness of voice. Note that for large $p$, $S$ will generally be very large. For instance, for $p=1$, i.e. if we were to require absolute certainty of a match, $S = \infty$.

\subsection{Pair match}
A \textit{pair match} is the event $\M$ such that a \textit{randomly drawn} pair of individuals, $a$ and $b$, are both assigned to the same bin.

If $a$ is assigned to a cell $c$, the probability that $b$ too is assigned to $c$ is simply $P(c) = \frac{1}{m}$.  Thus

\begin{equation}\label{eq:pair}
P(\M) = \frac{1}{m}
\end{equation}

A simple interpretation of $P(\M)$ is that we may expect to have to test 
\(
N(\M) = \frac{1}{P(\M)}
\)
random pairs of individuals to find a match. Ideally $N(\M)$ must be greater than the square of the population size to ensure that all voices are unique.

\subsection{Population match}
The population match problem is an instance of the famous birthday problem \cite{borja2007birthday}. Briefly, although there are 365 days in a year, in any group of only 23 people, there is a greater than 50\% probability that two of them will share a birthday. The apparent paradox is that although the number of people in the group is far smaller than the number of days of the year, the odds of finding a shared birthday are so high.

There is no real paradox, however. The high probability arises from the fact that we are considering only the \textit{existence} of a collision (of birthdays), without specifying the individuals. From a frequentist perspective, the interpretation would be that if we were to form many random groups of 23 people,  roughly half of them would have a pair of people who share a birthday.

In our context, the population match problem can be stated as follows: if we assign the $n$ individuals in our population independently to one of the $m$ voice cells, what is the probability that at least one of the cells has more than one individual in it?  The arithmetic is well known; nevertheless, we present it here for reference.

Let $T(m,n)$ be the \textit{total} number of ways in which $n$ individuals could be independently assigned to one of $m$ cells. Since each of the $n$ individuals could be assigned to any of the $m$ buckets, we have
\[
T(m,n) = m^n
\]

We want the probability of the event $\B$ that at least one cell has more than one individual in it.  However, it is simpler to calculate the probability of the complementary event $\bar{\B}$ that \textit{no} cell has more than one individual and subtract it from 1. Let $U(m,n)$ be the number of ways in which $n$ individuals can be assigned to $m$ voice cells so that no cell has more than one individual. This is simply the permutation
\[
U(m,n) = \frac{m!}{(m-n)!}
\]

The probability $P(\bar{\B})$ that no two individuals fall in the same voice cell is the ratio of $U(m,n)$ to $T(m,n)$: 
\begin{align}
P(\bar{\B}) &= \frac{U(m,n)}{T(m,n)} \\
&  = \frac{\frac{m!}{(m-n)!}}{m^n} \\
& = \frac{m \times (m-1) \times \cdots \times (m-n+1)}{m^n} \\
& = \prod_{i=1}^{n-1} \frac{m-i}{m} \\
&= \prod_{i=1}^{n-1} \left(1 - \frac{i}{m}\right)
\end{align}

Since $m$ and $n$ are large and the probability involves a large number of multiplications, this equation is better dealt with in the logarithmic domain where the multiplication becomes a summation:
\begin{equation}\label{eq:pp0}
\log P(\bar{\B}) = \sum_{i=1}^{n-1} \log \left(1 - \frac{i}{m}\right)
\end{equation}
$P(\B)$, the probability that at least two individuals fall within the same voice cell is the complement of this probability: 
\begin{equation}\label{eq:pp1}
P(\B) = 1 - P(\bar{\B})
\end{equation}

When $n < \sqrt{m}$, the $\frac{i}{m}$ terms in Equation \ref{eq:pp0} are small and it can be further simplified using the approximation $\log(1-x) \approx -x \quad \text{for~} x \ll 1$, giving us
\begin{align*}
\ln P(\bar{\B}) &= \sum_{i=1}^{n-1} -\frac{i}{m} \\
&= -\frac{1}{m}\sum_{i=1}^{n-1} i \\
&= -\frac{1}{m} \frac{(n-1)n}{2} \\
&= -\frac{n(n-1)}{2m}
\end{align*}

Exponentiating both sides of the above, we get
\[
P(\bar{\B}) = \text{e}^{-\frac{n(n-1)}{2m}}
\]
In the cases where $n^2 \ll m$ we can use the approximation \(e^{-x} \approx 1 -x \quad\text{for}~x \ll 1\) to get
\[
P(\bar{\B}) \approx 1-\frac{n(n-1)}{2m}
\]

The probability that at least two individuals have the same voice is the complement of this probability:
\begin{align}
P(\B) &= 1 - P(\bar{\B}) \\
&= 1 - \left(1 - \frac{n(n-1)}{2m}\right)
\end{align}
giving us
\begin{equation}\label{eq:prob}
P(\B) \approx \frac{n(n-1)}{2m}
\end{equation}

For large $n$, $(n-1) \approx n$, giving us
\begin{equation}\label{eq:prob2}
P(\B) \approx \frac{n^2}{2m}
\end{equation}
This result, fundamental to probability theory, scales quadratically with $ n $ and inversely with $ m $ \cite{Feller1968}.

\section{Computing actual probabilities}
With this in perspective, let us calculate these entities for the case where $n=10^{10}$ (10 billion), standing for the population of Earth. We quantize each of the 44 features into $q=10$ bins giving us $m=10^{44}$. 

We calculate exact match probability $P(\E)$ using Equation \ref{eq:exact01}, $N(\E)$, match at $p$ value $S$ for $p = 10^{-9}$ using Equation \ref{eq:atp1}, pair match probability $P(\M)$ using Equation \ref{eq:pair}, $N(\M)$, and the population match probability $P(\B)$ using Equation \ref{eq:prob}, and find them to be the following values.

\begin{align*}
P(\E) & = 1e-34 \\
N(\E) &= 1e+34 \\
S &= 1e+35\\
P(\M) &= 1e-44 \\
N(\M) & = 1e+44\\
P(\B) & = 5e-25
\end{align*}

As the numbers show, at this resolution of the acoustic features (10 levels per feature), finding a match for a given speaker,  finding a random pair of matching speakers, or even finding a population match are all virtually impossible.

However, $q=10$ is likely a level of resolution that is perceptually or computationally not achievable, so let's consider these probabilities at different quantization levels.

\subsection{A systematic calculation of chances for different range quantizations}

We now consider these metrics at different quantization levels for the features, for $q$ ranging from $2$ to $10$. For \textit{match at $p$} we have arbitrarily chosen $p = 10^{-9}$, representing a search through 1 billion groups before we find one that includes a match for the speaker.

\bigskip

\begin{table}[ht]
\centering
\caption{Collision metrics, for a population of $10^{10}$ individuals distributed over $q^{44}$ voice types.}%
\vspace{0.5em}
\label{tab:collision_probabilities_n10e10_k44}
\begin{tabular}{|>{\raggedright\arraybackslash}p{0.4cm}|
                >{\raggedright\arraybackslash}p{1.6cm}|
                >{\raggedright\arraybackslash}p{1.6cm}|
                >{\raggedright\arraybackslash}p{1.6cm}|
                >{\raggedright\arraybackslash}p{1.6cm}|}
\hline
\textbf{\( q\)} & \textbf{\(P(\mathcal{E})\)} & $S$ & $P(\mathcal{M})$ & $P(\mathcal{B})$ \\
\hline
10 & 1e-34 & 1e+35 & 1e-44 & 5e-25 \\
9  & 1.03e-32 & 9.7e+32 & 1.03e-42 & 5.16e-23 \\
8  & 1.84e-30 & 5.44e+30 & 1.84e-40 & 9.18e-21 \\
7  & 6.54e-28 & 1.53e+28 & 6.54e-38 & 3.27e-18 \\
6  & 5.77e-25 & 1.73e+25 & 5.77e-35 & 2.88e-15 \\
5  & 1.76e-21 & 5.68e+21 & 1.76e-31 & 8.80e-12 \\
4  & 3.23e-17 & 3.09e+17 & 3.23e-27 & 1.62e-07 \\
3  & 1.02e-11 & 9.84e+11 & 1.02e-21 & 0.0495 \\
2  & 5.68e-04 & 1.76e+04 & 5.68e-14 & 0.9999 \\
\hline
\end{tabular}
\end{table}

%

Table \ref{tab:collision_probabilities_n10e10_k44} has few surprises. At the highest levels of resolution ($q=10$), finding a match, either to a specific person ($P(\E)$, or even a random pair ($P(\M)$, is effectively impossible. It must be noted that $q=10$ represents fewer than 4 bits of resolution per feature, a not unreasonable ask.  At 3 bits of resolution, i.e. $q=8$, the probability of an exact match, $P(\E)$, remains at $\sim10^{-30}$, the size $S$ of the \textit{match at $p$} cohort required to find a match is over $10^{30}$ even for $p = 10^{-9}$, and the probability of a random pair matching, $P(\M)$, is $10^{-40}$. The population match probability $P(\B)$ of the mere existence of a match is $10^{-20}$.

Even at two bits of resolution or lower ($q=4$ or $q=3$), voices remain unique.  For $q=3$ the probability $P(\E)$ of an individual finding a match is $10^{-11}$, i.e. an order of magnitude worse than the inverse of the human population, and the size $S$ of the peer set needed to find a match with even a probability of $10^{-9}$ is two orders of magnitude greater than the population of the planet.

\begin{figure}[h]
\centering
\includegraphics[width=0.5\textwidth]{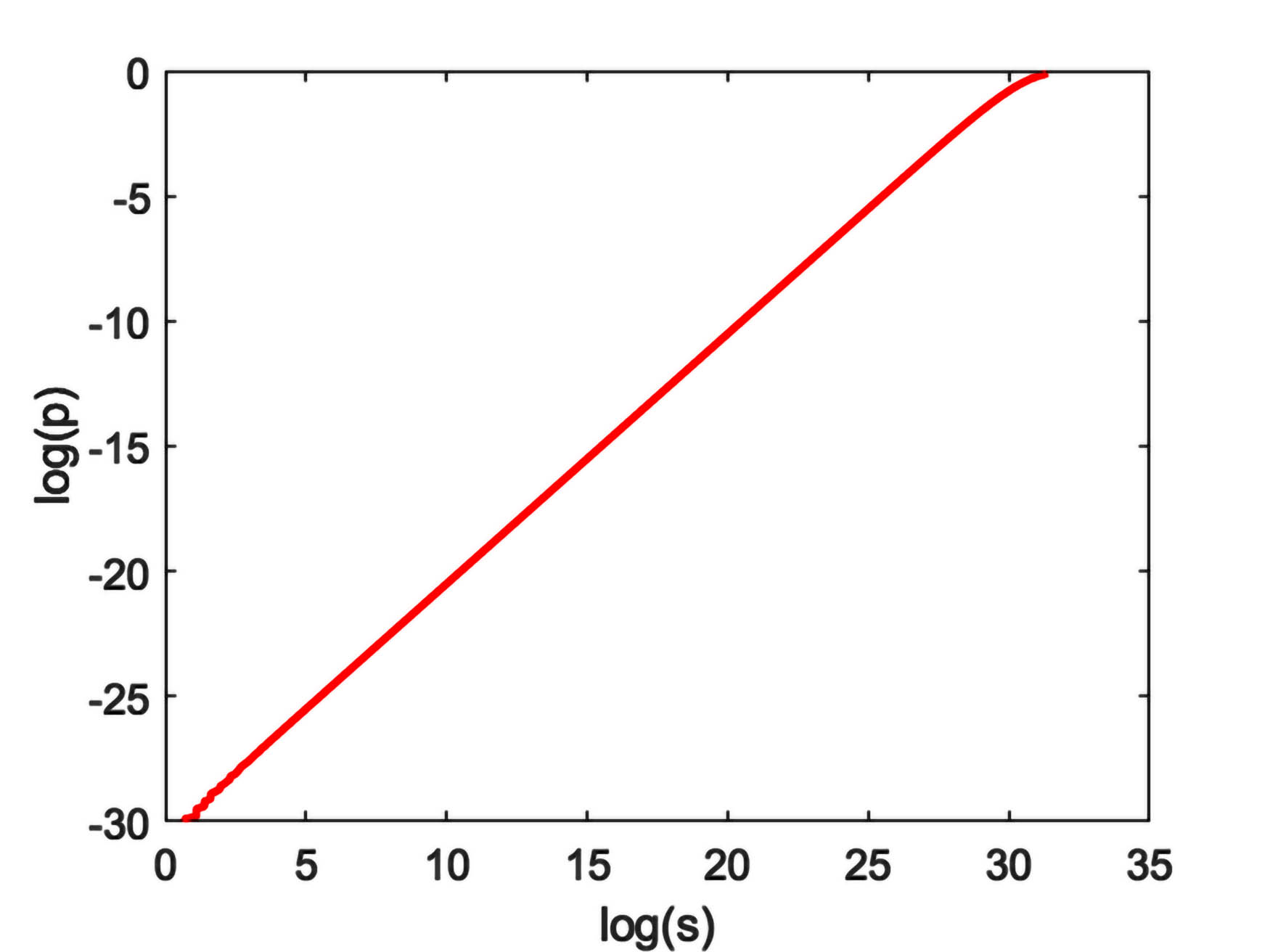}
\caption{A log-log plot of $p$-vs-peer set size $S$ for \textit{match at $p$} for $q=2$. The dependence of $s$ on $p$ is exponential.}\label{fig:pvss}
\end{figure}

It is only when we are down to one bit of resolution, i.e. $q=2$ that matches become more common.  Even here, the actual size $S$ of the peer set we must consider for any reasonable probability of match is non-trivial.  For probability $p=10^{-9}$ of finding a match, we must already consider 17600 peers.  For greater certainty this number grows exponentially with $p$ (linearly with $\log p$, as shown in Figure \ref{fig:pvss}). Thus, to have a probability of $10^{-6}$ of finding a match, one must consider a peer set of over 17 million peers.

The \textit{population match} metric, $P(\B)$, apparently provides a more sobering perspective. At two bits of resolution or less, the population match probability rises to ${\sim}10^{-7}$ for $q=4$,  ${\sim}0.05$ for $q=2$ and ${\sim}1.0$ for $q=2$. However, this figure too belies the true difficulty of obtaining a match. At $q=3$ the population match probability is approximately $0.05$, indicating that at a resolution of 3 levels there is a 5\% probability of the existence of individuals whose voices match. However, a better perspective is obtained when we realize that this result could be interpreted as follows: 
at two-bit resolution of the features, we could expect to have to redo the distribution of a population of $10^{10}$ individuals over one million times to find one in which a match occurs. Even at $q=3$ we would have to redo this over 20 times. It is only at a very coarse 1-bit resolution of features is finding a population match certain.
Interestingly, since we perceive many voices to be the same or similar, it appears that our auditory system perceives the equivalent of coarse partitions of the ranges of these signal characteristics implicitly.

\section{Conclusions}

While anthropologists have cataloged cranio-facial diversity and geneticists have mapped DNA variation globally, a global,  quantitative account that supports the uniqueness of human voice is missing and leaves a gap in the scientific literature. 

Much about the human voice and its range of variation is still unknown and uncharted. With sufficient, systematically collected voice data from across the world, it may be possible to gauge the approximate range of variation of each of the features listed in Table \ref{tab:independent-features}. We can then apply dimensionality reduction techniques to find the most significant quantizations for each of these entities. Such an exercise would give a more precise indication of the uniqueness of the human voice. For the purpose of this paper, however, the approximate ranges used in  Table \ref{tab:collision_probabilities_n10e10_k44} provide a good estimate.

It must be noted that the computations in Table \ref{tab:collision_probabilities_n10e10_k44} are only approximations, and neither represent upper bounds nor lower bounds. It is unclear at how much granularity each of the given 44 acoustic features contributes to the vocal identity of the speaker. On the other hand, voice production is a very complex process, and almost certainly includes additional variables not accounted for in the given 44 acoustic features; thus the actual individuality of voices may be greater than that quantified by Table \ref{tab:collision_probabilities_n10e10_k44}.

It is interesting to note that the calculations presented in Table \ref{tab:collision_probabilities_n10e10_k44}  can also guide some fundamental choices made in the analysis of voice signals for different voice technologies. Depending on the task they are applied to, voice technologies may need to generalize across the human population, or subdivisions of it, or be specific to each user. As examples of the former, we have automatic speech recognition (ASR) systems and voice translation systems. Since these systems cater to the canonical human voice, if the features in Table \ref{tab:independent-features} are used, their ranges can be quantized to two levels (low/high). This will help avoid over-fitting, and help generalize to the population. In contrast, for voice verification systems, the systems would need to maintain a resolution that allows them to differentiate between different voices with clarity. For these, it may be sufficient to quantize the feature ranges to three levels (low/medium/high). For human voice profiling systems, the resolution must be even higher, since subtle effects on voice due to the influence of myriad internal and external factors on the speaker (which may be time varying) must be captured and used to infer the speaker's bio-parameters. For these, depending on the number of bio-parameters deduced, we may need to quantize the feature ranges to more than three levels.

These calculations can also guide voice-cloning systems that use artificial intelligence techniques. In cases where explicit tokenizations based on the features in Table \ref{tab:independent-features} are provided to these systems, they may perform better if they are additionally constrained to lie within the feature range partitions that are characteristic of the individual voices they clone. This is likely to contribute to the quality of voices from a vocal production perspective --  it could bring them a step closer to generating voices that are indistinguishable from real human voices.

Finally, we recognize that there is an additional need to analyze special cases (siblings, twins, ethnically homogeneous cohorts) to show how anatomical and experiential covariance alters uniqueness guarantees. 

We close by noting that the methodology used to quantify uniqueness is not specific to voice, but may be applied to any biometric, where an independent set of features that characterize the biometric may be identified.

\end{document}